\documentclass[mathleft
]{an}
\usepackage{graphicx}
\usepackage{rotating}
\usepackage{lscape}
\usepackage{longtable}
\usepackage{times}
\usepackage{fge}
\begin{document}

% The following seven commands are intended for editorial usage and should be ignored by
% the author(s).
\Pagespan{789}{}% Document's page range.
% If second parameter is left empty, the last page is computed automatically.
\Yearpublication{2011}%
\Yearsubmission{2010}%
\Month{11}%
\Volume{999}%
\Issue{88}%
% \DOI{This.is/not.aDOI}%

\sloppy

\title{Interpretation of the historic Yemeni reports  
of supernova SN 1006: early discovery in mid-April 1006~?}

\author{R. Neuh\"auser\inst{1} \thanks{Corresponding author: \email{rne@astro.uni-jena.de}}
\and D.L. Neuh\"auser\inst{2}
\and W. Rada\inst{3}
\and J. Chapman\inst{4}
\and D. Luge\inst{1}
\and P. Kunitzsch\inst{5}
}

\titlerunning{SN 1006 from Yemen}
\authorrunning{Neuh\"auser et al.}

\institute{
Astrophysikalisches Institut und Universit\"ats-Sternwarte, Friedrich-Schiller Universit\"at Jena,
Schillerg\"a\ss chen 2-3, 07745 Jena, Germany (e-mail: rne@astro.uni-jena.de)
\and
Schillbachstrasse 42, 07743 Jena, Germany 
\and
Hilla University College, Babylon P.O.B. (386), Iraq (deceased August 2015)
\and
Center for East Asian Studies, Stanford University, 521 Memorial Way, Stanford, CA 94305-5001, United States
\and
LMU Munich, Germany (retired); home: Davidstrasse 17, 81927 M\"unchen, Germany
}

\received{2015}
\accepted{2016}
\publonline{ }


\abstract{The recently published Yemeni observing report about SN 1006 from al-Yam\={a}n{\={\i}}
clearly gives AD 1006 Apr $17 \pm 2$ (mid-Rajab 396h) as first observation date.
Since this is $\sim 1.5$ weeks earlier than the otherwise earliest reports (Apr 28 or 30)
as discussed so far, we were motivated to investigate an early sighting in more depth.
We searched for additional evidences from other areas like East Asia and Europe.
We found that the date given by al-Yam\={a}n{\={\i}} is fully consistent 
with other evidence, including:
(a) SN 1006 {\em rose several times half an hour after sunset} (al-Yam\={a}n{\={\i}}), 
which is correct for the location of \d{S}an$^{c}$\={a}' in Yemen for the time around Apr 17, 
but it would not be correct for late Apr or early May;
(b) the date ({\em 3rd year, 3rd lunar month, 28th day wuzi}, Ichidai Yoki) for an observation of a guest star in Japan 
is inconsistent (there is no day {\em wuzi} in that lunar month), 
but may be dated to Apr 16 by reading {\em wuwu} date rather than a {\em wuzi} date;
(c) there is observational 
evidence that SN 1006 was observed
in East Asia early 
or mid April; for the second half of April, a bad weather (early monsoon) period is not unlikely --
there is a lack of night reports; 
(d) the observer in St. Gallen reported to have seen SN 1006 {\em for three months}, 
which must have ended at the very latest on AD 1006 Jul 10, given his northern location, 
so that his observations probably started in April.
We conclude that the correctly reported details give quite high confidence in the fully
self-consistent report of al-Yam\={a}n{\={\i}}, 
so that the early discovery date should be considered seriously.
} 

\maketitle

\section{Introduction: SN 1006}

Historic observations of supernovae (SN) are essential to understand SNe, neutron stars, and SN remnants (SNR):
Historic reports can in principle deliver the date of the explosion 
(hence, the age of the SNR and, if existing, the neutron star)
together with a light curve (hence, possibly the SN type), 
sometimes the colour and its evolution, and the position of the SN,
which is needed to identify the SNR and, if existing, the neutron star. 
Such historic observations have been used very successfully for SNe 
1006 (from East Asia, Europe, and Arabia),
1054 (from East Asia and Arabia),
1181 (only from East Asia),
and SNe 1572 and 1604 (from East Asia and Europe),
plus a few more SNe from the 1st millenium AD (see Stephenson \& Green 2002, henceforth SG02, and references therein).
While the Arabic report of SN 1054 merely confirms a bright new star in Gemini/Taurus around AD 1054,
the reports of SN 1006 present a lot of detailed information (Goldstein 1965), 
which were used to identify the SNR (Gardner \& Milne 1965).

The transient celestial object of AD 1006 was listed as comet in Pingre (1783).
Humboldt (1851) lists it as {\em new star},
based on the St. Gallen chronicle, dated incorrectly to AD 1012, and placed incorrectly in Aries.
Sch\"onfeld (1891) corrects the date to AD 1006 
(consistent date shift in St. Gallen chronicle) and the location to Scorpius
(previous misreading of the Syriac {\em $^{c}$aqrab\={a}} for Scorpius as {\em 'emr\={a}} for Aries in Bar Hebraeus);
he already used the chronicle of Bar Hebraeus and its source,
namely the annals of Ibn al-Ath\={\i}r.\footnote{In the report by Lynn (1891),
an English summary of the paper by Sch\"onfeld (1891), which was written in German,
it is said that Sch\"onfeld (1891) would have corrected the
date from AD 1006 to 1012, but the opposite is true.}

Convincing evidence for SN 1006 was presented first by Goldstein (1965) based on the Arabic reports:
$^{c}$Al\={\i} ibn Ri\d{d}w\={a}n (lived from AD 988 or 998 until 1061 in Cairo, Egypt;
indeed, it was considered seriously before that he observed SN 1006 at an age of only eight years,
the calculations could have been done later) 
reported the ecliptic longitude (15th degree of Scorpius),
strong scintillation ({\em it twinkled very much ... large ... round in shape}),
the size and/or brightness ({\em 2.5 or 3 times as large as Venus ... the intensity 
of its light was a little more than the quarter of that of moonlight}),
the duration of the observations (some four months until conjunction with the Sun),
that it did not move relative to the stars 
({\em It remained where it was and it moved daily with its zodiacal sign})
and that he observed it as an eyewitness during 
{\em the beginning of my studies ... all I have mentioned is my own personal experience};
he listed the (calculated) positions of the planets as well as those of the Sun, 
the Moon, and its ascending node,
from which Goldstein (1965) deduced the date of his observations to be 
the evening of AD 1006 Apr 30, the earliest certain observation accepted in SG02.
$^{c}$Al\={\i} ibn Ri\d{d}w\={a}n also mentioned that 
{\em Sun and Moon met in the 15th degree of Taurus} when the transient object first appeared;
only from that statement we can conclude that it appeared close to the conjunction of moon and sun 
(new moon on AD 1006 Apr 30 at 9:08h UT), so that SN 1006 was probably sighted (by him) on the evening of Apr 30.

Goldstein (1965) presented another Arabic report of SN 1006 from 
Ibn al-Jawz{\={\i} (a historian, who lived AD 1116-1201 in Baghdad, Iraq) and,
based on Ibn al-Jawz{\={\i}, also by Ibn al-Ath\={\i}r 
(a historian, who was born AD 1160 in Jazirat Ibn Umar, 
now Cizre in Turkey, and who died in 1233 in Mosul, Iraq),
both about a very bright new star -- 
as well as a report from Morocco\footnote{By the author 
Ab\={u} l-\d{H}asan $^{c}$Al\={\i} b. $^{c}$Abdall\={a}h b. Ab\={\i} Zar$^{c}$ al-F\={a}s\={\i} 
(short: Ibn Ab\={\i} Zar$^{c}$, died in or after AD 1326) in the book entitled
{\em Al-An\={\i}s al-mu\d{t}rib bi-rau\d{d}at al-qir\d{t}\={a}s f\={\i} akhb\={a}r 
mul\={u}k al-maghrib wa-t\={a}r{\={\i}}kh mad\={\i}nat F\={a}s},
the mentioned town of {\em F\={a}s} is now called Fes in Morocco, 
an edition of the work appeared in 1972 in Rabat, Morocco.}
mentioning a {\em great star} [{\em najm}] {\em among the comets} and a {\em nayzak} ({\em spectacle} 
or {\em guest star} or {\em transient celestial object}, see Kunitzsch 1995).
While the reports of Ibn al-Jawz{\={\i} (and Ibn al-Ath\={\i}r) 
give the date of first appearance as Friday, the 1st day of the Muslim month
of Sha$^{c}$b\={a}n of the Muslim year 396h\footnote{The Islamic year 396 
hijra (396h) started 396 lunar years after the start of the lunar year
in which the Hijra took place, i.e. the emigration of the Islamic Prophet Mu\d{h}ammad from Mecca to Medina, 
known as Hijra; this era, i.e. the year 1h started in the evening of AD 622 Jul 16
(1st day from evening of 16th to evening of 17th July)
according to most scholars -- but it may have been one day earlier
(evening of 15th to evening of 16th July),
see, e.g., de Blois (2000); according to Gautschy (2011, 2014),
see www.gautschy.ch/$\sim$rita/archast/mond/Babylonerste.txt,
new moon was on AD 622 Jul 14 (Julian calendar), so that the crescent new moon was
not visible in the evening of AD 622 Jul 14,
hardly visible in the evening of AD 622 Jul 15, 
but it was well visible in Mecca, Saudi Arabia, in the evening of AD 622 Jul 16 (Neugebauer 1929).
While the date of the 1st day of year 1h was mistakenly given one day too early in 
footnote 1 in Rada \& Neuh\"auser (2015), all other dates there are correct.
The uncertainty in the exact date of the start of the Hijra era
affects only the conversion between the calculated Muslim calendar and some other calendar (like the Julian or Gregorian calender),
and this uncertainty amounts to only one day.} (and
visibility until Dh\={u} al-Qa$^{\rm c}$dah, i.e. roughly three months), 
the Moroccan report mentions that {\em it began to appear in the beginning (i.e. 1st) of Sha$^{c}$b\={a}n}
and that {\em it lasted for a period of six months} (Goldstein 1965), 
i.e. from early May 1006 (Sha$^{c}$b\={a}n 396h) until October; 
however, SN 1006 was in conjunction with the Sun in October,
so that they could have observed it until at most the heliacal setting in the middle of September;
it is more likely that this {\em period of six months} were meant as rough statement
(like about half a year) than that they observed heliacal rising in November.

The Islamic date of 1 Sha$^{c}$b\={a}n 396h corresponds to AD 1006 May 2/3, 
evening to evening (e.g. Goldstein 1965), but only in the calculated Islamic calendar,
while the real start of a month was fixed by observations (not by a calculated calendar);
Muslim dates run from one evening to the next evening, 
a month starts with the evening of the first sighting of the crescent of a new moon.
It was confirmed in Rada \& Neuh\"auser (2015) 
that the conversion of 1 Sha$^{c}$b\={a}n 396h to AD 1006 May 2/3 is
correct when considering the first observation of the crescent (and also regarding the given week-day).
In general, unless more information is available, 
the conversion from the calculated Islamic calendar to the Julian or Gregorian
calender has an uncertainty of some 2 days due to (a) uncertainty in the start of the Hijra era (one day),
(b) uncertainty as to which months and years in history had one extra day (in addition to 354 days in 
12 lunar months -- given that a synodic month is not exactly 29.5 days), and (c) uncertainty as to when
the new crescent moon was sighted first
(e.g. Spuler \& Mayr 1961, Spuler 1963, Neuh\"auser \& Kunitzsch 2014).
Goldstein (1965) also gives the Arabic texts for the four Arabic reports presented.

Goldstein (1965) also gives an English translation 
of a Syriac report of SN 1006 by Bar Hebraeus (born AD 1226 in Malatya in Turkey, 
died 1286 in Maragha, now Iran), where it was specified that the new star was observed in the zodiacal sign of Scorpius.
Cook (1999) presented another Arabic report 
of SN 1006 from Ya\d{h}y\={a} ibn Sa$^{c}$\={\i}d al-An\d{t}\={a}k\={\i}, Patriarch of Antioch
(now Antakya, Turkey), who extended the chronicle of 
Eutychius of Alexandria (Egypt) for the time since circa AD 939 and died in AD 1066, 
according to which the new star was seen for four months
since {\em Saturday, 2nd day in Sha$^{c}$b\={a}n} of the year 396h (AD 1006 May 3/4).
Most recently, Neuh\"auser, Ehrig-Eggert, and Kunitzsch (2016) presented another 
original report about SN 1006, namely written by Ibn S{\={\i}}n\={a} (Avicenna). 

From the ecliptic longitude of the SN as given by $^{\rm c}$Al\={\i} ibn Ri\d{d}w\={a}n
(and an error bar from $^{\rm c}$Al\={\i} ibn Ri\d{d}w\={a}n's assumed measurement precision)
together with the declination limit from a St. Gallen observation 
of this SN (Goldstein 1965) and the Chinese right ascension range
(from the Chinese {\em lunar lodge}), it was then possible to constrain the location of the SN and to identify the SNR
(Gardner \& Milne 1965).

SN 1006 is believed to have taken place on AD 1006 Apr 30 or earlier
(see Rada \& Neuh\"auser 2015) in the constellation of Lupus 
with the following basic parameters: 
\begin{itemize}
\item distance being $2.18 \pm 0.08$ kpc from the proper motion of ejecta in mas/yr and the shock velocity
of filaments in km/s (Winkler et al. 2003),\footnote{The distance determination by Jiang \& Zhao (2007), 
$\sim 1.56$ kpc, is highly uncertain: it was obtained by interpreting a presumable observation of SN 1006 
in AD 1016 as an effect of re-brightening of parts of the SNR, 
while SG02 argued that this late observation date is a mistake in a historical document.} 
\item extinction being A$_{\rm V} = 0.32 \pm 0.03$ mag (Schaefer 1996 from various techniques) or
A$_{\rm V} = 0.31 \pm 0.10$ mag (Winkler et al. 2003) from the reddening of the very blue sub-dwarf
(Schweizer \& Middleditch 1980) showing strong, broad absorption lines due to the SNR, 
so that it is located in the background at 1.05-2.1 kpc (Burleigh et al. 2000) 
to 1.5-3.3 kpc (Schweizer \& Middleditch 1980), 
\item peak apparent brightness being $-7.5 \pm 0.4$ mag from distance for a SN type Ia (Winkler et al. 2003), and 
\item the apparently young SNR G327.6+14.6 was identified as its remnant 
(Gardner \& Milne 1965, Reynolds et al. 1994). 
\end{itemize}
While Damon et al. (1995) and Firestone (2014) claim that a $^{14}$C signal from SN 1006 was observed (in AD 1009),
Menjo et al. (2005) argued that the $^{14}$C amplitude around AD 1009 may be consistent with typical
Schwabe cycle modulation. (A $^{14}$C detection three years after the SN might be possible due to the carbon
cycle, which takes a few years.) 

\section{Arabic text(s) of SN 1006 from Yemen}

Most recently, Rada \& Neuh\"auser (2015) presented two more Arabic reports of SN 1006 in
both Arabic and English translation, namely by the Yemeni historians al-Yam\={a}n{\={\i}} and Ibn al-Dayba$^{c}$.
An English translation of the latter text was first presented in SG02 quoting private communication with one of us (WR).
In Rada \& Neuh\"auser (2015), the dating of the observations is discussed in detail.

The date given by both al-Yam\={a}n{\={\i}} and Ibn al-Dayba$^{c}$, mid-Rajab meaning 15th of Rajab, 
converts in the calculated Islamic calendar to AD 1006 Apr $17 \pm 2$. 
The dating uncertainty arises only from the uncertainties in the conversion from the Islamic to the Julian calendar 
and the observation of the crescent new moon.
The additional information from al-Yam\={a}n{\={\i}}, that the new star was rising
half an hour after sunset, is best fullfilled on Apr 17.
While the full moon rose before SN 1006 on Apr 16, which could have made
an observation of SN 1006 more difficult, the moon rose after SN 1006 since Apr 17, as seen from Yemen.
See Table 1 in Sect. 3.3.
On the other hand, the Islamic observer in Yemen probably observed on Apr 16 
(full moon was on Apr 16 at 9:35h UT) in the
evening in order to check whether it is full moon;
until the 14th of the lunar month, the date is specified in historic Arabic texts by giving the number of days since the
last crescent new moon, then the 15th of the month is specified as {\em mid} of the month,
and then, since the 16th of the month, the date is specified
by giving the number of days or nights expected until the next crescent new moon, 
assuming that the month has 30 days (de Blois 2000).
The middle day of a month must not neccessarily be the full moon day, because the 
Arabic lunar month does not start with new moon, 
but with the first observaion of the crescent; however, if this particular month lasted
30 days (for the observer in Yemen), 
the date of mid-Rajab would correspond to Apr 16 evening to Apr 17 evening.
SN 1006 was probably already observable on Apr 16 given its separation from 
the moon -- of course depending on its brightness that evening.

While the other Arabic and East Asian observations all were obtained between 
the geographic latitudes of $30^{\circ}$ and $35^{\circ}$ north, 
we discuss here in detail this additional observation 
from \d{S}an$^{c}$\={a}', the capital of Yemen, i.e. at $15.3^{\circ}$ north. 
We repeat briefly the English translations of the Arabic texts (Sect. 2).
In Sect. 3, we discuss in detail the evidence for an early discovery in mid (or even early) April.
We conclude our findings in Sect. 4.

If the early observation of SN 1006 in mid-April in Yemen can be shown to be plausible,
this could have important consequences for the light curve and SN type.

The first (earlier) text is from the book entitled
{\em Bahjat al-zaman f\={\i} t\={a}r{\={\i}}kh al-Yaman} 
written by al-Yam\={a}n{\={\i}} (died AD 1342, more details about him in Rada \& Neuh\"auser 2015);
the edition of al-Hubaishi \& al-Sanab\={a}ni (1988) was used.
The Arabic text is shown in figure 1 in Rada \& Neuh\"auser (2015).

We repeat the English translation here
(with an unlikely text variant given in square brackets): 
\begin{quotation}
In the night of mid-Rajab (or: 15th of Rajab), in the year 396h,
a star appeared from the east at half an hour after sunset.
It was four times as large as Venus.
[It was as large as Venus and rose several times after sunset.]
It was not circular, but nearer to an oblong.
At its ends, there were lines like fingers.
It showed a great turbulence as though it was seen in disturbed water.
Its light rays were similar to sunlight.
It appeared in the zodiacal sign [{\em burj}] of Libra in Scorpius
and remained unchanged like that.
In the night of mid-Rama\d{d}\={a}n, its light started to decrease and gradually faded away.
\end{quotation}

The 2nd (later) text is from the book entitled
{\em Kit\={a}b Qurrat al-$^{c}$uy\={u}n f\={\i} akhb\={a}r al-Yaman al-maim\={u}n} about the history of Yemen,
written by Ibn al-Dayba$^{c}$ (AD 1461 - 1537, more details about him in Rada \& Neuh\"auser 2015).
Rada \& Neuh\"auser (2015) used manuscript number 416 from the Wadod Center for indexing and edited books;
this manuscript is a copy written in AD 1680; see figures 2 \& 3 in Rada \& Neuh\"auser (2015) for the Arabic text. 
There is also an edition of Ibn al-Dayba$^{c}$'s work {\em Qurrat al-$^{c}$uy\={u}n} by al-Akwaa' al-Hiwali 
as publication of the Hiw\={a}li Yamani Library (\d{S}an$^{c}$\={a}', Yemen).

The text by Ibn al-Dayba$^{c}$ (first presented in SG02) is clearly derived from al-Yam\={a}n{\={\i}}
as already shown in Rada \& Neuh\"auser (2015). 
It does not add any additional information.
In the following interpretation, we will consider mainly the al-Yam\={a}n{\={\i}} text.

The last sentence of al-Yam\={a}n{\={\i}} ({\em In the night of mid-Rama\d{d}\={a}n, its light started
to decrease and gradually faded away}),
and also even in Ibn al-Dayba$^{c}$ ({\em its light diminished and it gradually faded away}), both for mid June,
does not say that the brightness dropped suddenly -- as was reported by $^{c}$Al\={\i} ibn Ri\d{d}w\={a}n:
{\em it ceased all of a sudden}
at a time between mid-Aug and mid-Sep 1006, when the Sun was in sextile with the new star.\footnote{In Rada
\& Neuh\"auser (2015), {\em July} was given for the end of the observation by $^{c}$Al\={\i} ibn Ri\d{d}w\={a}n
as mistake.}
It does not mean that the Yemeni observations were restricted to those two months from mid-April to mid-June,
but rather that SN 1006 remained roughly constant until mid-June, and then started to get fainter.

\section{Discussion of the early observation}

It would be quite surprising if the observer, who is the original source for the reports of the Yemeni authors,
really has observed and detected SN 1006 already several days before the 
other Arabic observers (e.g. Apr 30 by $^{c}$Al\={\i} ibn Ri\d{d}w\={a}n), 
and all East Asian observers, so that this early date was rejected by SG02 as artificial
(based only on the text by Ibn al-Dayba$^{c}$).

We would now like to discuss several arguments which can be interpreted in favour of an early observation,
at least not excluding an early observation:

\subsection{Yemen: {\em half an hour after sunset}}

The text of al-Yam\={a}n{\={\i}} says: 
\begin{quotation}
In the night of mid-Rajab (or: 15th of Rajab), in the year 396h,
a star appeared from the east at half an hour after sunset.
It was four times as large as Venus.
\end{quotation}
The position of SN 1006 indeed did rise {\em half an hour after sunset} at the location of \d{S}an$^{c}$\={a}'
in the middle of April 1006, while this statement would not be true at the end of April or early May.
We cannot exclude that al-Yam\={a}n{\={\i}} (and Ibn al-Dayba$^{c}$ or their source) calculated much later
that SN 1006 rose half an hour after sunset at the given night of mid-Rajab 396h (AD 1006 Apr 17),
which may not be impossible (as $^{c}$Al\={\i} ibn Ri\d{d}w\={a}n had probably also calculated the
positions of Sun, Moon, and planets as given in his report several decades after SN 1006, Goldstein 1965, SG02),
but this would appear more doubtful. There are otherwise no obviously calculated facts in the report.
The report by our most original and earliest Yemeni source, al-Yam\={a}n{\={\i}}, is self-consistent
regarding the early date and rising time. We should therefore consider it seriously.

\subsection{Southern location of \d{S}an$^{c}$\={a}'}

Since \d{S}an$^{c}$\={a}' is at 2400 m sea level and since it has a 
clear horizon towards the south,\footnote{The Jabal al-Nabi Shu$^{c}$ayb,
the highest mountain in Arabia with 3666 m, is due west-south-west 
from \d{S}an$^{c}$\={a}', while SN 1006 was rising in the
evening in the south-east.} an observation from here earlier than all other known observations may not appear impossible:
going from such a large height to roughly sea level,
where most of the other, later observers were located (e.g. Cairo, Japan, China) can change the
atmospheric extinction for object low on the horizon by some 4 mag (Schaefer 1993);
the limit for serendipitous discovery of a new star on the sky by naked-eye is
some 0 to 2 mag according to Clark \& Stephenson (1977) and Strom (1994).

This consideration does not exclude that other observers at low altitude 
(e.g. in China or Japan,
see Sect. 3.4)
would have observed SN 1006 in mid April, as they could have observed later in the night,
when SN 1006 was higher above the horizon. The professional Chinese and Japanese astronomers
have observed all night.
The Yemeni observers, though, may have observed mainly around and shortly after sunset,
close to the time of the last prayers.

\subsection{Rising of SN 1006 before the Moon on Apr 17 only from Yemen}

It is well possible that the Yemeni (and other Arabic) observers checked for the
full moon on the evenings of Apr 15, 16, and 17 (full moon on Apr 16 at 9:35h UT),
in order to know the date in their lunar month relative to the full moon.

While on AD 1006 Apr 16, the Moon was above the horizon earlier than the location of SN 1006
as seen from Yemen, the Moon rose half an hour after SN 1006 on Apr 17 and even later on Apr 18,
so that an observation of SN 1006 may appear more probable on Apr 17 from Yemen.
At the other relevant observing sites (Morocco, Iraq, Japan,
and in particular also in Cairo, Egypt, and Kaifeng, China), the Moon was rising before
SN 1006 even on Apr 17, e.g. 20 minutes before the SN 1006 as seen from Kaifeng, 
today's name of the capital of China at that time. 

\begin{table*}
\caption{Times of apparent rising of SN 1006, sunset, and moonrise for
\d{S}an$^{c}$\={a}', Yemen and Kaifeng, China (times given in UT).
We also list the separation ({\em sep}) between SN 1006 and the moon for the dates given,
as seen from \d{S}an$^{c}$\={a}' (left) and Kaifeng (right) -- always given for two hours after
local moonrise, so that both the SN and the moon were visible.
Bold face times indicate cases, where SN 1006 rose before the Moon
(e.g. on Apr 17 in \d{S}an$^{c}$\={a}', but not in Kaifeng.} 
\begin{tabular}{lllll|llll} \hline
1006 & \multicolumn{4}{c}{for \d{S}an$^{c}$\={a}', Yemen} & \multicolumn{4}{c}{for Kaifeng, China} \\
Apr  & SN rise & sunset    & moonrise    & sep SN/Moon & SN rise        & sunset& moonrise    & sep SN/Moon \\ \hline
15   & 15:54   & 15:16     & 14:34       & $26.2^{\circ}$ & 12:34       & 10:57 & 10:05       & $27.6^{\circ}$ \\
16   & 15:50   & 15:17     & 15:27       & $20.4^{\circ}$ & 12:30       & 10:57 & 11:05       & $20.7^{\circ}$ \\
17   & {\bf 15:46} & 15:17 & {\bf 16:20} & $22^{\circ}$ & 12:26         & 10:58 & 12:06       & $20.8^{\circ}$ \\
18   & {\bf 15:42} & 15:17 & {\bf 17:17} & $30.1^{\circ}$ & {\bf 12:22} & 12:59 & {\bf 13:07} & $23.3^{\circ}$ \\
19   & {\bf 15:38} & 15:17 & {\bf 18:14} & $41.4^{\circ}$ & {\bf 12:18} & 12:59 & {\bf 14:07} & $38.9^{\circ}$ \\
20   & {\bf 15:34} & 15:17 & {\bf 19:10} & $54.4^{\circ}$ & {\bf 12:14} & 11:00 & {\bf 15:07} & $51.9^{\circ}$ \\ \hline 
\end{tabular} 
\end{table*}

This consideration does not exclude that other observers further north (e.g. in China or Japan)
would have observed SN 1006 in mid April close to full moon, because the moon was sufficiently
well separated from SN 1006, so that SN 1006 may have been observable -- depending on its brightness.
The Yemeni observers had a particulary good reason (full moon) to observe in mid April.

\subsection{Possible observation in Japan on Apr 16 or 28}

In the medieval Japanese chronicle Ichidai Yoki, an independant and original source
based on {\em Abe Yoshimasa, teacher in astronomy}, we can read: 
\begin{quotation}
[AD 1006 Apr 28:] ... in 3rd year, 3rd lunar month, 28th day wuzi [25], a guest star entered Qi,
\end{quotation}
(Stephenson et al. 1977).
The date given as {\em 3rd year, 3rd lunar month, 28th day} corresponds to AD 1006 Apr 28.
East Asian reports often specify the date in addition with the day count
in the sexagenary system of numbering days continuously from 1 to 60.
However, as discussed in SG02, there is no day called {\em wuzi} (25) 
in the third lunar month of that year; the name and number of the day AD 1006 Apr 28
is {\em gengwu} (7); the relevant characters ({\em wu} and {\em zi} compared to {\em geng} and {\em wu})
are very similar, so that already Kanda (1935) suggested that this guest star was
indeed observed on the 28th day of that lunar month,\footnote{Note that the Chinese 
and Japanese started the day-count in each (lunar) month
with what we call {\em new moon}, i.e. conjunction of moon and sun, as confirmed by the fact that all of the dates of
solar eclipses from (at least) AD 700 to 1200 are dated to {\em the first day of the month}, see listing in Xu et al. (2000),
while the Arabs started the lunar month day count with the first sighting of the crescent (Quran, Sura 2, 189).}
i.e. already on AD 1006 Apr 28
(and that, later on, a scribe made a mistake with the date number).
However, as pointed out by SG02, two separate scribal errors need to be assumed.
The {\em 28th day} of that lunar month (i.e. 1006 Apr) would definitely be a few days before
the Chinese ({\em 2nd day of the fourth lunar month}, i.e. May 2/3)
and Arabian observations,
the latter being {\em at the beginning of Sha$^{c}$b\={a}n}, 
i.e. beginning of May, and 1006 Apr 30 for $^{c}$Al\={\i} ibn Ri\d{d}w\={a}n. 

It is very well possible that the latter character in the sexagenary date alone was mistranscribed.  
If Kanda (1935) is correct regarding his emendation of the second character in the sexagenary date
(from {\em zi} to {\em wu}), 
but the first character ({\em wu}) is left to stand as it is in the received text, 
the date becomes a {\em wuwu} day (55), i.e. AD 1006 Apr 16.  
{\em Wu} might easily be mistranscribed as {\em zi}, 
as both characters contain two horizontal and one vertical strokes; 
the two characters are differentiated by an initial curved stroke in {\em wu} and hooks at the end of 
the first and third stroke of {\em zi}. 
See Fig. 1.

\begin{figure}
\begin{center}
{\includegraphics[angle=0,width=5cm]{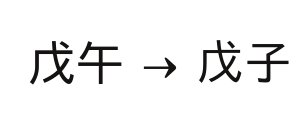}}
\end{center}
\caption{{\bf Chinese characters:} If the Chinese characters have unintentionally changed
due to a mistake made by a copying scribe
from those shown in the left ({\em wuwu}, i.e. day 55) to those shown in the right ({\em wuzi}), 
the original date of the observation could have been AD 1006 Apr 16.}
\end{figure}

The Apr 16 date is then not the 28th day of the lunar month, as also specified in the source.
However, such dates often consist only of the year of the emperor, the number of the lunar month,
and the date in the sexagenary system, leaving out the day within the lunar month (see, e.g., Sect. 3.5).
It is possible that the original source did not contain such a lunar day, but that it was amended later,
or that 16 was mistranscribed as 28.
If we posit a {\em wuwu} day (Apr 16), 
then the Ichidai Yoki record corresponds very closely to that of the Yemeni observers.  
However, just as the sexagenary date casts doubt on the Apr 28 date,
the lunar date casts doubt on the possible Apr 16 date.
The dating of the event in the Ichidai Yoki remains uncertain.
There are two other reports from Japan on this SN: Meigetsuki (13th century) and 
Dainihonshi (completed AD 1715) both list a report for May 1, see e.g. SG02,
but it is obvious that the latter depends on the former.
The report discussed here from Ichidai Yoki is an independent medieval chronicle of unknown date.

According to SG02, the observation of SN 1006 within a few days at different places in Arabia 
(30 Apr to May 2 only) 
as well as in China and Japan (about 1 May) may provide evidence against an earlier observation elsewhere,
we have to see that there were 8 to 6 days from the earliest previously accepted
first observation (28 Apr in Japan or 30 Apr in Cairo) to the latest reported first observation
(May 6 in China: {\em 3rd year, 4th lunar month, day wuyin [=AD 1006 May 6]. A Zhoubo star appeared ...} 
from SG02 from Wenxiao Tongkao), which is quite a long time for such a
bright SN, in particular also for an observation around the new moon.

A possibly relatively long time between the first observations in the different countries
(it may have been even more days between the first detections in different countries)
can also be seen as evidence for unstable weather at least in some of those 
places. E.g., Rada \& Neuh\"auser (2015) provided evidence for bad weather on
AD 1006 May 1 in Antiochia, now Turkey, and maybe Mosul, Iraq: The reports from there 
mention explicitly {\em Saturday, the 2nd of Sha$^{c}$b\={a}n} 
and {\em Friday, the 1st of Sha$^{c}$b\={a}n}, respectively,
so that both started the month of Sha$^{c}$b\={a}n on the evening of (our) Thursday, May 2,
even though the crescent new moon would have been well visible at those sites on
the evening of May 1. The non-detection of the crescent on May 1 may indicate bad weather.

\subsection{Possible early observation on Apr 3 in the SE in China}

Reports of an even earlier sighting run as follows: \\
(i) Wenxian Tongkao: {\em Jingde reign period, third year, third lunar month (day) yisi [42] (=AD 1006 Apr 3). 
A guest star [ke xing] appeared (chu) in the south-east direction}, \\
(ii) Songshi Annals: {\em Jingde reign period, third year, third lunar month (day) yisi [42] (=AD 1006 Apr 3). 
A guest star [ke xing] appeared at the south-east}, and \\
(iii) Songshi Astronomical treatise: {\em Jingde reign period, 
third year, third lunar month (day) yisi [42] (=AD 1006 Apr 3).
A guest star [ke xing] appeared at the south-east}, \\
the complete citations from SG02 with their additions in round brackets and
our additions in square brackets.

This guest star may have been another object, e.g., a comet (SG02).
There are, however, also a few arguments in favour of a possible interpretation of this guest star as SN 1006: \\
(a) The guest star was seen in the {\em south-east} like SN 1006.  \\
(b) The more general word for {\em guest star} [ke xing] was used and not a more specific word
for {\em broom star} or {\em tailed star} [hui xing] or {\em fuzzy star} [xing bo],
which would have indicated a comet. \\
(c) There are no additional Chinese or other records available on any additional comet or other object
in or around early April.  \\
(d) An important political meeting on April 17 is reported, 
which could have been a consequence of the very bright magnitude of the new star:
{\em On the jiwei day (Apr 17), admonitory ministers were summoned to court and asked to speak openly about
what should and should not be done} (Song shi 7.130),
but the reason for the meeting is not specified.

That the information from China about this SN is sparse, in particular for April and May,
may be due to the difficult interpretation at that time: A solar eclipse was expected
for 1006 May 30, which would have to be interpreted in a more negative sense for the
emperor; the new star became a Zhoubo star, for which the historic Chinese texts
offered both positive or negative interpretations (SG02). Once it became obvious that
the solar eclipse did not take place at the capital, it was not necessary any more
to consider a negative interpretation (for both eclipse and the new star), so that
one could opt for the positive interpretation of the new Zhoubo star,
which of course met the approval of the emperor. This is fully consistent
with the Chinese texts dated to May 30, see SG02,
and it could possibly explain why Chinese sources are unusually quiet
about the bright new star in its first few weeks.

\subsection{No other East Asian observations Apr 17-28/30 (and Apr 4-15)}

If the Japanese have observed SN 1006 on Apr 16
(and maybe the Chinese already on Apr 3),
then again later since May 1, 
it would be surprising that there are no reports left from
the professional astronomers in China and Japan about
any observations inbetween, i.e. from Apr 17 to the end of April (or even Apr 4 to 15).

Are there any East Asian observations known for the intermediate periods from Apr 17 to 
the end of April (or even from Apr 4 to 15)?
Are there any East Asia night reports before 1006 Apr 16, where no guest star is mentioned~?
There is only one Chinese observation known for April 1006, namely for AD 1006 Apr 14 reporting: 
\begin{quotation}
Emperor Zhenzong of Song, 3rd year of the Jingde reign period, 3rd month, day bingchen (53).
In the north a scarlet vapour extending across the sky (and a white vapour penetrated the Moon),
\end{quotation}
citing from Xu et al. (2000), a slightly different translation
in Keimatsu (1975), both from Songshi 60.1308, without the text in brackets also in Yau et al. (1995).
This is a probable aurora according to the criteria given in Neuh\"auser \& Neuh\"auser (2015),
namely northern directions, aurora-typical colour, and night-time (implicitly given with the 
moon).\footnote{
In the list of candidate aurorae in Hayakawa et al. (2015), this event is listed twice,
once with {\em 1006 Apr 14 R[ed] V[apour] n[orth] [in] Kaifeng [moon phase] 0.46} (near full moon)
and once with {\em 1006 Apr 14 W[hite] V[apour] near the moon [in] Kaifeng [moon phase] 0.46} (near full moon);
one of the two texts is from the astronomical treatise ({\em Tianwen zhi}) to the {\em Songshi}, 
the other from its treatise on general omenology ({\em Wuxing zhi}).
In the same list of candidate aurorae in Hayakawa et al. (2015), there is an additional entry:
{\em 1006 May [without day] Y[ellow] V[apour] near the moon [in] Kaifeng}, also from Songshi;
the translation of this entry is: {\em On this date yellow vapour like a pillar penetrated the moon}.
The date for this event is uncertain; however, as it is given as the {\em guimao} (40) day in the fourth month, 
when in fact there was no {\em guimao} day in the fourth month; a {\em guimao} day did occur at the beginning 
of the fifth month (1006 May 31) and at the beginning of the third month (1006 Apr 1) --
both, however, so close to new moon that the text ({\em penetrated the moon}) does not fit
to the given sexagenary date. There is another instance of the same phrase ({\em Yellow vapour like a pillar 
penetrated the moon}) dated to the 4th month of the 3rd year of the Tianxi reign period (AD 1019) 
given without the {\em guimao} date, though there is a guimao date in that month;
it is possible that the event somehow got transposed to the wrong reign period;
Hayakawa et al. (2015) list {\em Y[ellow] V[apour]} for 1019 May 8, but by mistake omitted here
{\em near the moon}, which is clearly given in the original Chinese; they give 0.04 as moon phase
(new moon May 7), so that again the text ({\em near the moon}) is not consistent with the moon phase
for this date; though the entry does not actually specify which day within the lunar month the event occured on,
the {\em guimao} day (40) in that lunar month was AD 1019 May 23, i.e. close to
full moon (May 21/22), when a lunar halo display would be possible.
In any case, this event is not an aurora, but more likely a lunar halo pillar.
}
What is reported as {\em a white vapour} penetrating {\em the Moon} may well be some halo effect 
around the Moon, which is well possible two days before full moon;
for a discussion of the aurora sighting around full moon, see Chapman et al. (2015). 

There are indeed no additional East Asian observations known for the remaining time of AD 1006 Apr 17
until the end of April. 

There is evidence for the fact that relevant Chinese documents are missing:
\begin{quotation}
... on the 2nd day of the 4th lunar month the Zhoubo star was seen. The official astronomer reported
it immediately. The [Song] Shilu [for] the Xiangfu reign period, 9th year, 4th lunar month, [day] gengchen
should be consulted for further details,
\end{quotation}
quoting SG02 from Xu Zizhi Tongjian Changbian -- however, the mentioned Song Shilu is lost (SG02).

Furthermore, there is also evidence that Zhou Keming, a prominent astronomer in China,
was on a mission
in April 1006, so that he may not have been able to consult documents 
(for the interpretation) on SN 1006 early. 
In the Biography of Zhou Keming (AD 954-1017), we can read: 
\begin{quotation}
During the 3rd year of the Jingde reign period, a large star appeared in the sky at the west of Di.
No-one could determine (its significance) ... At the time, (Zhou) Keming was away on a mission to Lingnan,
On his return, he urgently requested to reply ... He said: "I have checked the (astrological manuals)
Tianwen Lu and the Jingzhou Zhan ... the star is known by the name Zhoubo, which is yellow in colour
and really brilliant in its light. The country where it is visible will prosper greatly ..."
The Emperor approved and acceded to his request. He then promoted him to the post of Librarian
and Escort of the Crown Prince,
\end{quotation}
cited from SG02 with their additions in brackets. 

We can see that Zhou Keming was on a mission to Lingnan 
(southern China, Goldstein and Ho Peng Yoke 1965),
while the guest star first appeared,
that no one present could (or was allowed to) interpret its astrological meaning,
and that -- upon his return -- he checked the old documents 
about the astrological meaning of the bright guest star,
identified it as Zhoubo star, and reported his interpretation to the Emperor.

Other historical documents specify that the Emperor was informed on AD 1006 May 30: 
\begin{quotation}
[AD 1006 May 30:] The Director of the Astronomical Bureau reported that previously,
on the 2nd day of the 4th lunar month [May 1], during the initial watch of the night,
a large star had been seen. Its colour was yellow ... According to the star manuals,
there are four categories of auspicious stars. One of them is called Zhoubo; its colour
is yellow and it is really brilliant; it presages great prosperity to the state over
which it appears ... The officials congratulated the Emperor,
\end{quotation}
from SG02 from Song Huiyao Jigao. The astronomer, who is informing the
Emperor here, is also called {\em superintendent astronomer} in the same document.

Furthermore, we can read:
\begin{quotation}
Jingde reign period, 3rd year [AD 1006-1007], there was a large star seen in the sky ...
Zhou Keming, the chief official of the Spring Academy reported that according to the Tianwen Lu
and the Jingzhou Zhan, the star was a Zhoubo,
\end{quotation}
from SG02 from Shaofu Yitang Qinghua, also quoted in Goldstein and Ho Peng Yoke (1965),
who point out that the {\em Tianwen Lu} and {\em Jingzhou Zhan} are lost.

That the information from China about this SN is sparse, in particular for April and May,
may be due to the difficult interpretation at that time, i.e. until after the
expected solar eclipse at the end of May, as mentioned in Sect. 3.6.

\subsection{Possibly bad weather (monsoon) in East Asia}

The East Asian monsoon affects large parts of China, Korea, and Japan; 
the onset of the summer monsoon with pre-monsoonal rain
over South China is typically in early May, but can also start a few days or weeks earlier;
the summer monsoon with many rainy phases starts in the South China Sea 
and then moves northward to Japan (June) and Korea (July).
There are no East Asian night-time observations known at all between AD 1006 Apr 17 and the end of April.
For AD 1006 Apr 14, we have evidence for a halo display in the south as seen from China (see above: {\em white
vapour penetrated the Moon}).
The lack of reports for the time AD 1006 Apr 17 until the end of April may be due to 
either bad weather or the fact that the reports were lost.

\subsection{SN 1006 observed in St. Gallen for 3 months}

A monk from St. Gallen, Switzerland, reported for AD 1006: 
\begin{quotation}
Nova stella apparuit insolitae magnitudinis, aspectu fulgurans, et oculos verberans,
non sine terrore. Quae mirum in modum aliquando contractior, aliquando diffusior, et iam extinguebatur interdum.
Visa est autem per tres menses in intimis finibus austri, ultra omnia signa quae videntur in coelo,
\end{quotation}
cited after Pertz (1826) from the Annales Sangallenses maiores (covering AD 709-1056);
its second part (AD 919-1056) was written by different authors, Hepidannus being one of them,
a St. Gallen monk, who lived in the 2nd half of the 11th century and died AD 1088, 
i.e. not necessarily an eyewitness of SN 1006 himself.
The above text was translated as follows: 
\begin{quotation}
[AD] 1006. A new star of unusual size appeared, it was glittering (fulgurans)
in appearance and dazzling (verberans) the eyes, causing alarm. 
In a wonderful manner it was contracted, sometimes spread out, and moreover sometimes
extinguished. It was seen, nevertheless, for three months in the inmost limits of the
south, beyond all the constellations which are seen in the sky,
\end{quotation}
citing Stephenson et al. (1977) and SG02 with their brackets and additions;
for what they translate as {\em constellations}, the Latin has {\em signa},
which can mean {\em signs} or {\em zodiacal signs}.

\begin{figure*}
\begin{center}
{\includegraphics[angle=270,width=17cm]{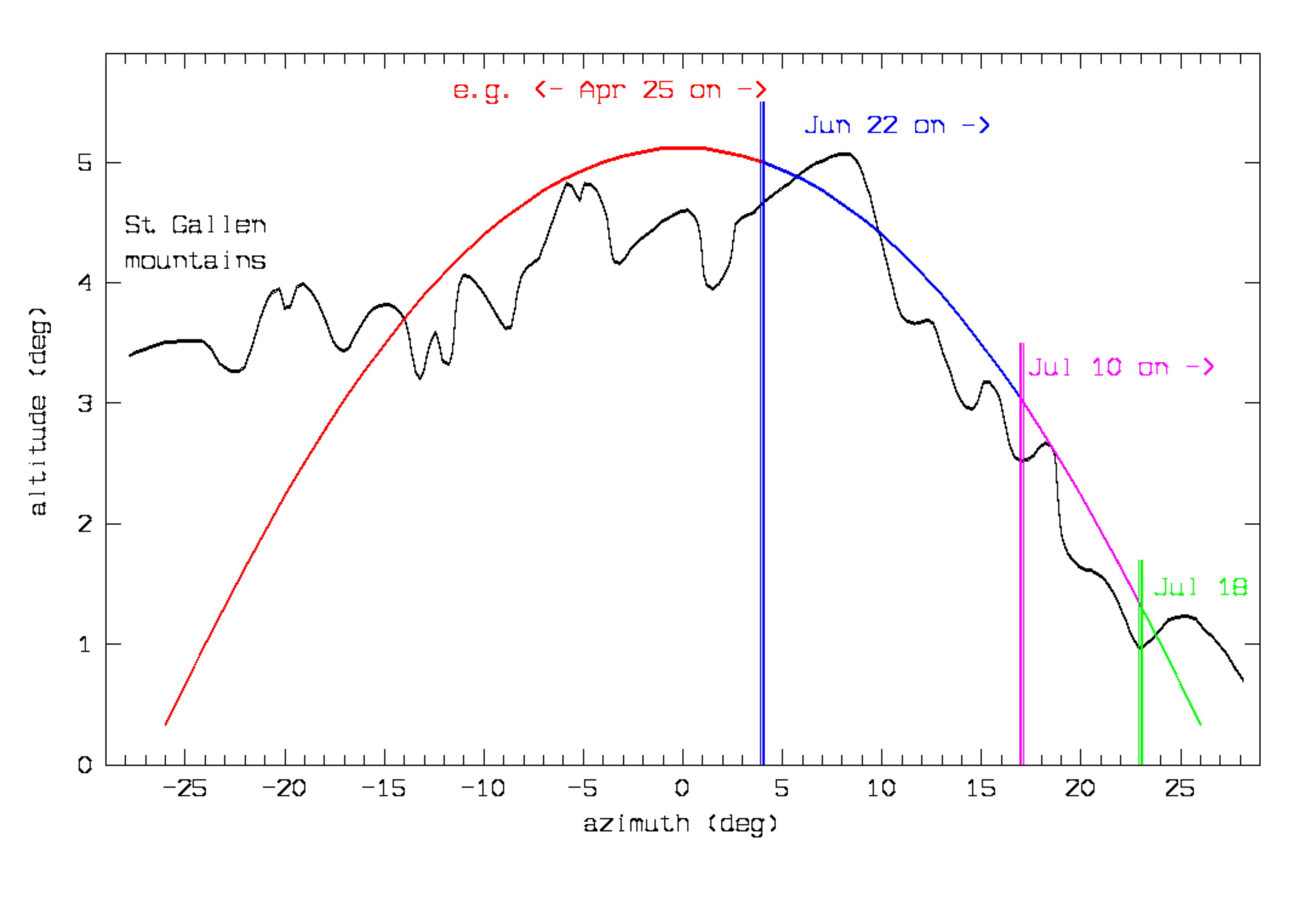}}
\end{center}
\caption{{\bf Visibility of SN 1006 from St. Gallen:}
The altitude (in degrees) of (the position of) SN 1006 is plotted versus the azimuth (in degrees) for St. Gallen. 
The black line shows the mountain top as seen towards the south from St. Gallen monastery (700 m high)
according to Stephenson et al. (1977) with Mount S\"antis as highest peak at 2503 meter.
The additional curve shows the path of the position of SN 1006 as seen from St. Gallen.
Since Jul 18, the position of SN 1006 would be seen after sunset only as plotted (in green) 
to the right of the (rightmost, green) line,
i.e. not visible any more above the horizon or the mountains;
since Jul 10, the position of SN 1006 is seen after sunset only as plotted (in pink or green) 
to the right of the (pink) line;
since Jun 22, the position of SN 1006 is seen after sunset only as plotted (in blue or pink or green) 
to the right of the (blue) line
(i.e. visible for all azimuths $\ge 4^{\circ}$);
and since Apr 25, the position of SN 1006 is seen after sunset as plotted (in red or blue or 
pink or green), i.e. it was above the mountain
(except of course behind the mountain top at azimuth $\sim 6-10^{\circ}$ for a brief period).
If SN 1006 was seen in St. Gallen {\em for three months}, i.e. at least for 2.5 months, 
and if it was visible last around Jul 10 (for about one minute after sunset) or earlier, 
then the first observation should have been in April.
Given that the observer described the star to be {\em sometimes extinguished}, 
he must have been at an altitude such that the star, within the night, was sometimes seen above the mountain
and sometimes being briefly occulted by the mountain top,
so that the star was seen between $\sim 4^{\circ}$ to $\sim 5^{\circ}$ above a perfect flat horizon
(given the height of Mount S\"antis).
Therefore, the altitude of the observer was somewhere between 700 m and $\sim 1100$ m.
}
\end{figure*}

The relevant part about the length of the observation of SN 1006 is 
{\em Visa est autem per tres menses}, i.e. {\em per tres menses},
which clearly means {\em for three months} or {\em throughout three months}, 
e.g. from the beginning of some month 1 (not neccessarily a calendar  month) until the end of month 3.
With {\em for three months}, the author of this part of the St. Gallen annales 
did not necessarily mean {\em three full months}, 
he may have rounded down or up (i.e. 2.5 to 3.5 months).
The wording clearly does not mean {\em in three (different, subsequent calendar) months},
which could then have meant that it was observed first at the end of month 1 (e.g. May) 
and last at the beginning of month 3 (e.g. July).
It is noteworthy to mention that the St. Gallen chronicle does not mention the
duration of visibility just in passing, but explicitely ({\em nevertheless, for three months}),
in spite of the difficult conditions (high mountains, strong extinction).

The southern horizon as seen from St. Gallen has high mountains with Mount S\"antis
straight towards the south being the highest one with 2503 m, located 20 km south of St. Gallen.
The monastery is at an elevation of some 700 m, but the monks may have observed from a slightly
higher point nearby;
e.g. somewhat closer to the mountain.
The highest point in today's St. Gallen is 1074 m. 
The summit of Mount S\"antis as seen from either the monastery or the
higher point is only $\sim 4^{\circ}$ to $\sim 5^{\circ}$ above a perfectly flat mountain-less horizon.
While the monk may in principle have observed from an even higher point,
since the location itself is not specified in the text, the range in degrees given above,
i.e. only about one degree fron $\sim 4^{\circ}$ to $\sim 5^{\circ}$ above horizon,
must indeed be as small as given: the text specifies that 
{\em it} [SN 1006] ... {\em moreover sometimes extinguished.}
The observer tells us that SN 1006 was sometimes seen and sometimes {\em extinguished},
which is well possible given the mountain range, where higher parts sometimes block the star light.
Hence, this statement limits the range in altitude and, therefore, also the range in
the height of the observing location, whereever it was (even if outside the monastery).
See Fig. 2.

If the star observed in St. Gallen was indeed SN 1006, 
then SN 1006 ($\delta _{1006} = -37^{\circ} 34^{\prime}$) 
was only up to $5^{\circ}$ above a perfectly flat (mountain-less)
horizon at its location ($47^{\circ}25^{\prime}$ north). 
However, the horizon was furthermore limited by mountains (SG02):
At an eastern azimuth, the true horizon due to mountains 
barely allowed celestial observations below $4^{\circ}$ above
the perfect flat horizon, while at a western azimuth of $\ge 10^{\circ}$, 
celestial objects $\le 3^{\circ}$ above the perfect flat horizon were visible (Fig. 2). 
(If the observer went to a place higher up than the monastery,
then SN 1006 could be seen a bit better and maybe a bit longer,
but one criterion of seeing it for the last time 
only in the very last minute after sunset is already very hard.)

We can now estimate the time of the year 
when SN 1006 was visible from St. Gallen above the mountains, see Fig. 2.

Let us first estimate the last observing date:
Given that there were no day-time observations reported for SN 1006 
(except the report from Morocco: {\em its first appearance was before sunset}),
we can assume that SN 1006 was visible only after local sunset in St. Gallen.
Due to its very low altitude and strong atmospheric extinction as seen from St. Gallen,
a day-time observaion of SN 1006 from St. Gallen is much less likely than from any other place,
where SN 1006 was seen. 
For SN 1006 being $3^{\circ}$ 
above the perfect flat horizon (but less than 1 degree above the mountains), 
it was last visible from St. Gallen on AD 1006 July 10 at 
an extinction corrected apparent magnitude of about $-1 \pm 1$ mag; 
and on June 22 with $\sim -2 \pm 1$ mag 
for about one minute after sunset for $5^{\circ}$ above  
the perfectly flat horizon (but less than one degree above the mountain); see Fig. 2;
we have neglected refraction here, which would amount to less than $1^{\circ}$.

If it was last visible on or around July 10 (or earlier), when was it first sighted? 
As mentioned above, with {\em for three months}, the author(s) of the St. Gallen annales 
means at least 2.5 months, namely until July 10 (or earlier), see above.
Then, he would have started to have seen SN 1006 on or around April 25 or earlier.
On and around April 25, SN 1006 would have been $\ge 3^{\circ}$ 
above a perfectly flat St. Gallen horizon 
(and about one degree above the mountains) for quite some time after sunset.

There is no regular weather pattern (like a monsoon in East Asia) in St. Gallen and central Europe 
as a whole in late April or early May
(except maybe that the weather changes a lot in Europe in April), it could have been clear on many evenings.
That not many other observers have noticed SN 1006 in Europe can be due to its extreme southerly declination,
so that only very experienced and educated scholars (like monks) would detect it.

Also the Annales Beneventani (southern Italy, $6^{\circ}$ south of St. Gallen)
report about a new bright star in 1006 and
use the wording {\em per tres menses} ({\em for three months}): \\
AD 1006: {\em Clarissima stella effulsit, et siccitas magna per tres menses fuit}, \\
which we translate as follows: \\
{\em A very brilliant star shone, and a large drought happened for three months}, 
(also given in SG02).\footnote{Additional European sightings do not mention the
date or length of the observations (SG02).}  \\
It is quite likely that the two items reported, a new bright star and a three-month drought,
are meant to be connected. Given that this observer located $6^{\circ}$ south of St. Gallen
and that he does not have high mountains towards the south,
he should have been able to observe SN 1006 for longer than in St. Gallen --
and indeed, while the St. Gallen report ({\em for three months}) can mean
at least 2.5 months, the Beneventani report ({\em for three months} can mean up
to 3.5 months, both may be rounded, they are not inconsistent with slightly different time spans.

A few more European annals mention a {\em cometes} for AD 1006,
namely Li\'ege and Lobbes, Belgium, also Venice, Italy, as well as Metz and Mousson, France (SG02).
Some of them are further north than St. Gallen, but probably just report what they heard
from St. Gallen. That they use the Latin word {\em cometes} (usually translated as {\em comet})
should not worry us, because at that time it meant {\em transient celestial object},
like the Arabic {\em nayzak}. The annals from St. Gallen and Benevento, though, do not
use the word {\em cometes} for 1006 indicating that the observers there noticed that this
transient celestial object was different from what we today call a {\em comet} --
indeed, a new/very brilliant star.

\section{Summary}

We have discussed the Arabic texts of the observation of SN 1006 
by al-Yam\={a}n{\={\i}} and Ibn al-Dayba$^{c}$ from Yemen,
also in comparison with other Arabic, East Asian, and European observations, in particular
in regard to the early sighting around AD 1006 Apr 16 and 17.

The relevant information from al-Yam\={a}n{\={\i}}, the more original text, is as follows:
\begin{itemize}
\item {\em In the night of mid-Rajab, in the year 396h, a star appeared ...}, 
i.e. possibly already in the evening of AD 1006 Apr 17 $\pm 2$,
\item {\em a star appeared from the east at half an hour after sunset}, 
which is consistent with AD 1006 Apr 17 evening,
\end{itemize}
These two statements are fully consistent with each other: 
only on and around 17 April 1006 (mid-Rajab 396h), 
SN 1006 rises half an hour after sunset as seen from Yemen.

SG02 rejected such an early sighting, but based only on the derived variant from Ibn al-Dayba$^{c}$: \\
{\em In the year 396h, in the night of mid-Rajab,
a star like Venus appeared.
It regularly rose half an hour after sunset.} \\
Here, the additional information about the rising time 30 min after sunset is corrupt
(it did not rise {\em regularly} half an hour after sunset).

SN 1006 was discovered in Arabia and Asia around Apr 30/May 1 
(but maybe even around Apr 3 in China and Apr 16/17 in Japan and Yemen),
all these dates are around either new moon or full moon.
The observations may have been facilitated by the 
observation or search for the moon phase 
in order to know the relative date within the lunar month.
Societies with a lunar calendar perform more celestial observations around new and full moons.

Also, a somewhat rare opportunity to observe at the same time during the night the four planets
Venus, Mars, Jupiter, and Saturn (together even with the Moon until May 17) started just on AD 1006 May 1, 
so that additional observations in the first hours of the nights may have started around May 1, 
even though close conjunctions of three planets did not happen, 
which were otherwise often reported by the Chinese.
Then, since May 22, Mercury was also seen together
with the four other naked-eye planets at the start of the nights,
namely until the end of June (May 31 to June 15 also with the Moon). 
For AD 1006 Aug 5, it was noticed that {\em Mercury, Jupiter, and Venus met in Liu},
i.e. in a lunar mansion (Xu et al. 2000).
In addition, Mars was stationary in June 1006, 
so that it may have been observed more closely also before. 
SN 1006 was seen as an additional bright object in the first hours of all those nights.

We presented the following evidence in favour of an early observation (earlier than the end of April): 
\begin{itemize}
\item The report that {\em a star appeared from the east at half an hour after sunset} (mid-Rajab 396h)
is fully consistent with AD 1006 Apr $17 \pm 2$, 
but it would not be consistent with late April or early May.
\item \d{S}an$^{c}$\={a}', Yemen, is quite high ($\sim 2400$ m) and far south,
both facilitating an early observation.
\item The Yemeni observer may have undertaken observations of the moon phase 
since around AD 1006 Apr 16 (full moon) in order to know the relative date within the lunar month;
SN 1006 was near the full moon in the south-east.
The other Arabic observers, who observed SN 1006 first at the end of April or early May,
were searching for the crescent new moon, in order to start a new month.
\item A guest star was seen in Japan possibly already on Apr 28 -- 
or even on Apr 16 (we suggest this alternative possibility for the Japanese text).
\item A guest star was seen in China on Apr 3 in the SE, i.e. the correct direction for SN 1006;
neither a tail nor motion relative to the stars were mentioned.

\item There are no reports about any other East Asian observations known for the period from Apr 17
until the end of April, possibly due to lost documents or bad weather.
\item It may be that SN 1006 was not observable in East Asia in the 2nd half of April 1006 due to early monsoon.
\item SN 1006 was observed in St. Gallen, Switzerland, {\em for three months} and was last visible there 
$3^{\circ}$ (or $5^{\circ}$, respectively) above the horizon AD 1006 July 10 (June 22, respectively) 
for about one minute after sunset,
so that the first observation should have been in April.
\end{itemize}

We found multiple evidence for an early observation in mid April:
a new star on April $17 \pm 2$ in Yemen,  
detection of a new star in St. Gallen already in April,
a new star on April 16 (or 28) in Japan, and
a possible observation on Apr 3 in China.
That there are not more records could be due to bad weather or lost document,
or because the Chinese had problems with the interpretation given the expected
solar eclipse for end of May.
It could be that more records will be found: 
$^{c}$Al\={\i} ibn Ri\d{d}w\={a}n wrote of SN 1006 that {\em other scholars from time to time have followed it}.

\acknowledgements
We acknowledge the Moon phase predictions by 
Rita Gautschy on www.gautschy.ch/$\sim$rita/archast/ mond/Babylonerste.txt.
WR would like to thank Sahi Hassoun al-Ta`i of Hilla University College for obtaining 
the second manuscript from the Wadod Center;
we would like to acknowledge the Wadod Center for indexing and editing books, 
which was established in memory of Ms. Sheikha al-Murry.
RN also thanks the Institut f\"ur Geschichte der Arabisch-Islamischen Wissenschaften, Frankfurt,
where he consulted the al-Hubaishi \& al-Sanab\={a}ni edition of al-Yam\={a}n{\={\i}}'s
work {\em Bahjat az-zaman f\={\i} t\={a}r{\={\i}}kh al-Yaman} as well as the auto-biography of Ibn al-Dayba$^{c}$.
RN acknowledges Frank Giessler for providing an electronic data file for the altitude of the St. Gallen
mountain tops (from Stephenson et al. 1977) for plotting Fig. 2.
We would like to thank an anonymous referee for good suggestions,
also for focusing our discussion.

{}


\begin{thebibliography}{}

\bibitem{} al-Hubaishi, A. \& al-Sanab\={a}ni, M.A. (Eds.), 1988, 
al-Yam\={a}n{\={\i}}: Bahjat az-zaman f\={\i} t\={a}r{\={\i}}kh al-Yaman,
Dar al-Hikma al-Yamania, \d{S}an$^{c}$\={a}', Yemen

\bibitem{} Brecher, B.R., Lieber, E., Lieber, A.E., 1978, Nat, 273, 728

\bibitem{} Burleigh, M.R., Heber, U., O'Donoghue, D., Barstow, M.A., 2000, A\&A, 356, 585

\bibitem{} Chapman, J., Neuh\"auser, D.L., Neuh\"auser, R. Csikszentmihalyi, M., 2015, AN, 336, 530

\bibitem{} Clark, D.H. \& Stephenson, F.R., 1977, The Historical Supernovae, Pergamon

\bibitem{} Cook, D., 1999, JHA, 30, 131

\bibitem{} Damon, P.E., Kocharov, G.E., Peristykh, A.N., Mikheeva, I.B., Dai, K.M., 1995, CRC 2
(24th International Cosmic Ray Conference, Vol. 2), 311

\bibitem{} de Blois, F.C., 2000, {\em T\={a}r}\textit{\={\i}}{\em kh}, in: Bosworth, C.E., van Donzel, E.,
Heinrichs, W.P., Lecomte, G. (Eds.) Encyclopaedia of Islam, new edition, Vol. X, Leiden: Brill

\bibitem{} Dreyer, J.L.E. 1906, History of the planetary system from Thales to Kepler, Oxford Univ. Press

\bibitem{} Firestone, R.B., 2014, ApJ, 789, 29

\bibitem{} Gardner, F.F. \& Milne, D.K., 1965, AJ, 70, 754

\bibitem{} Gautschy, R., 2011, Zeitschrift f\"ur \"Agyptische Sprache und Altertumskunde 178, 1

\bibitem{} Gautschy, R., 2014, JHA, 45, 79

\bibitem{} Goldstein, B.R., 1965, AJ, 70, 105

\bibitem{} Goldstein, B.R. \& Ho Peng Yoke, 1965, AJ, 70, 748

\bibitem{} Gonzalez-Hernandez, J.I., Ruiz-Lapuente, P., Tabernero, H.M., Montes, D., Canal, R., Mencez, J., Bedin, L.R., 2012, Nat, 489, 533

\bibitem{} Green, D.A., 2009, Bull. Astron. Soc. India, 37, 45

\bibitem{} Hamuy, M., Phillips, M.M., Suntzeff, N.B., Schommer, R.A., Maza, J., Smith, R.C., Lira, P., Aviles, R., 1986, AJ, 112, 2438

\bibitem{} Hartner, W., 1965, Isis, 56, 438

\bibitem{} Hayakawa, H., Tamazawa, H., Kawamura, A.D., Isobe, H., 2015, Earth, Planets and Space, 67, 82

\bibitem{} Ho Peng Yoke, 1962, Vistas, 5, 172

\bibitem{} Humboldt, von, A., 1851, Kosmos, Vol. 3

\bibitem{} Jiang, S.Y. \& Zhao, F.Y., 2007, Chin. J. Astron. Astrophys., 7, 325

\bibitem{} Kanda, S., 1935, Nihon Temmon Shiryo, Koseisha, Tokyo

\bibitem{} Keimatsu, M., 1975, A chronology of aurorae and sunspots in China, Korea, and Japan, Part VI (AD 1000-1200),
Ann. Sci. Kanazawa Univ. Report 12

\bibitem{} Kunitzsch, P., 1991, al-Man\={a}zil, in: Bosworth, C.E., van Donzel, E.,
Heinrichs, W.P., Lecomte, G. (Eds.), Encyclopaedia of Islam, new edition, Vol. VI, Leiden: Brill

\bibitem{} Kunitzsch, P., 1995, al-Nudj\={u}m, in: Bosworth, C.E., van Donzel, E.,
Heinrichs, W.P., Lecomte, G. (Eds.), Encyclopaedia of Islam, new edition, Vol. VIII, Leiden: Brill 

\bibitem{} Lynn, W.T., 1891, Observatory, 14, 265

\bibitem{} Menjo, H., Miyahara, H., Kuwana, K., Masuda, K., Muraki, Y., 
Nakamura, T., 2005, Proc. 29th Internat. Cosmic Ray Conf. Pune, 2, 357-360

\bibitem{} Morrison, L. \& Stephenson, F.R., 2004, JHA, 35, 327

\bibitem{} Neugebauer, P.V., 1929, Astronomische Chronologie, Berlin and Leipzig

\bibitem{} Neuh\"auser, R. \& Kunitzsch, P., 2014, AN, 335, 968

\bibitem{} Neuh\"auser, R. \& Neuh\"auser, D.L., 2015, AN, 336, 225, arXiv:1503.0158

\bibitem{} Neuh\"auser, R., Ehrig-Eggert, C., Kunitzsch, P., 2016, AN in press, arXiv:1604.03798

\bibitem{} Payne-Gaposchkin, C., 1956, Introduction to astronomy, Eyre and Spottiswoode, London

\bibitem{} Pertz, G.H. (Ed.), 1826, Monumenta Germaniae Historica, Vol. 1, reprinted Stuttgart 1963

\bibitem{} Pingre, A.G., 1783, Cometographie, Traite Historique et Theorique des Cometes, Vol. 1, Paris

\bibitem{} Pingree, D., 1986, Picatrix, The Warburg Institute, U of London

\bibitem{} Rada, W. \& Neuh\"auser, R., 2015, AN, 336, 249

\bibitem{} Reynolds, S.P., Lyutikov, M., Blandford, R.D., Seward, F.D., 1994, MNRAS, 271, L1

\bibitem{} Said, S.S., Stephenson, F.R., Rada, W., 1989, Records of solar eclipses in Arabic chronicles, in:
Bull. School of Oriental and African Studies, U London, Vol. LII Part I

\bibitem{} Schaefer, B.E., 1993, Vistas, 36, 311

\bibitem{} Schaefer, B.E., 1996, ApJ, 459, 438

\bibitem{} Sch\"onfeld, E., 1891, AN, 127, 153

\bibitem{} Schweizer, F. \& Middleditch, J., 1980, ApJ, 241, 1039

\bibitem{} Sezgin, F., 1971, Geschichte des arabischen Schrifttums, Vol. IV, Brill, Leiden

\bibitem{} Spuler, B., 1963, Der Islam. Zeitschrift f\"ur Geschichte und Kultur des Islamischen Orients, 38, 154

\bibitem{} Spuler, B. \& Mayr, J., 1961, W\"ustenfeld-Mahler'sche 
Vergleichungs-Tabellen, Dt. Morgenl\"andische Ges., Steiner Wiesbaden

\bibitem{} Stephenson, F.R. \& Green, D.A., 2002, {\it Historical Supernovae and Their Remnants}, Oxford, Clarendon (SG02)

\bibitem{} Stephenson, F.R., Clark, D.H., Crawford, D.F., 1977, MNRAS, 180, 567

\bibitem{} Strom, R.G., 1994, A\&A, 288, L1

\bibitem{} Wheeler, J.G., 1990, Supernovae, World Scientific, Singapore

\bibitem{} Willingale, R., West, R.G., Pye, J.P., Steward, G.C., 1996, MNRAS, 278, 740

\bibitem{} Winkler, P.F., Gupta, G., Long, K.S., 2003, ApJ, 585, 324

\bibitem{} Xu, Z., Pankenier, D.W., Jiang, Y., 2000, East Asian archaeoastronomy, Gordon and Breach

\bibitem{} Yau, K., Stephenson, F., Willis, D., 1995, 
A catalogue of auroral observations from China, Korea, and Japan (193 BC - AD 1770),
Technical report, Rutherford Appleton Lab

\end{thebibliography}
\end{document}